# INTRODUCING "OPTO": PORTAL FOR OPTICAL COMMUNITIES IN INDONESIA

Tomi Budi Waluyo and Laksana Tri Handoko

Research Centre for Physics – Indonesian Institute of Sciences
Kompleks PUSPIPTEK, Serpong 15314, Indonesia
Tel.: (021) 7560570, Fax.: (021) 7560554, E-mail: tomibudiwaluyo@lipi.fisika.net

*Abstract–* Since January 1, 2005 we launched "OPTO" Portal, a website dedicated to optical communities in Indonesia. The address of this portal is http://www.opto.lipi.go.id and is self-supporting managed and not for commercial purposes. Our aims in launching this portal are to benefit Internet facility in increasing the communities' scientific activity; to provide an online reference in Indonesian language for optics-based science and technology subjects; as well as to pioneer the communities' online activities with real impacts and benefits for our society. We will describe in the paper the features of this portal that can be utilized by all individuals or members of optical communities to store and share information and to build networks or partnership as well. We realized that this portal is still not popular and most of our aims are still not reached. This conference should be a good place for all of us to collaborate to properly utilize this portal for the advantages to the optical communities in Indonesia and our society at large.

## I. INTRODUCTION

In scientific communities, portals become highly desired as scientific research moves toward utilizing interdisciplinary research requiring access to non traditional data sets and tools such as the Internet [1]. As defined by the Wikipedia [2], portal is a website that functions as a point of access to information on the Internet from diverse sources in a unified way. It is also interesting to note David Morrison's definition which distinguishes a portal from a simple HTML page with framesets by saying that "a portal is an application or device that provides a personalized and adaptive interface for people to discover, track and interact with other relevant people, applications and content" [3]. Since portal provides an environment that simplifies access to a variety of resources and services to address the needs and goals of the community, we are developing a portal for optical communities in Indonesia. We called it the Opto Portal and its address is http://www.opto.lipi.go.id which is self-supporting managed and not for commercial purposes. We describe in this paper the features of this Portal that can be utilized by all individuals or members of optical communities to store, share, and access information as well as to build networks or partnership, and its progress so far since it was launched in January 1, 2005.

## II. FEATURES

This Portal is built based on open source applications (such as PERL) and, since it is dedicated to the Indonesian people especially for high school and undergraduate students as an on line source for optics-based science and technology subjects, its contents is in Indonesian language. We offer in this Portal features like mailing list, discussion forum, articles, agenda, database, search engine, etc. Figure 1 shows one of the pages of this Portal.

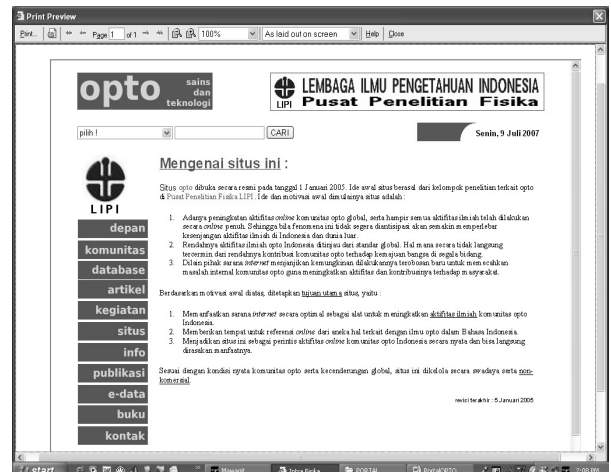

Figure 1. "About Us" page of the OPTO Portal.

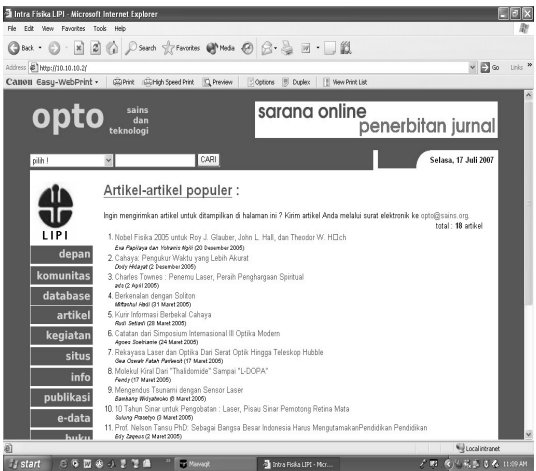

Figure 2. Contents of the "Article" page.

We design the features of this Portal as follows:

- User-friendly and full web interfaces for easy maintenance in all aspects and modules. No technical requirement for maintainers, except surfing the web.
- Multi maintainers with multiple access lists and multi level privileges.
- Automatic processes in all aspects to reduce maintenance works.
- Several modules to fit the needs of each scientific communities :
    - Posting regular articles.
    - Open forum.
    - Favorite related links.
    - Related scientific journal links.
    - Mailing-list.
    - Sharing electronic data.
    - Related books.
    - Related agenda.
    - Member database.
    - Storing non-journal publication (theses, etc).
    - Searchable data index for all modules.
- Search-engines is friendly designed to enhance access rate from global visitors.

To illustrate features offered in this Portal, Figure 2 shows the list of articles written by contributors (can be sent via E-mail to opto@sains.org) that can be accessed by all visitors.

### III. PROGRESS SO FAR

Since its activation, as always happened in most popular websites, the Opto Portal has also been suffered from the crackers/hackers. Nowadays, the crackers are getting improved and their techniques are quite advanced. In order to overcome that life-time problems, we keep updating the kernels to the latest one, although it couldn't anticipate the new attacks, especially the DOS (denial-of-service) type attacks that are the most difficult one. It sends huge traffics to the servers that is unfortunately undistinguishable from the regular traffics. The second type of attacks are the automatic and indiscriminate postings to several forms available in the Portal.

For the DOS attacks, we have limited the traffic flow from anonymous sources in a certain period. This technique could successfully reduce the inappropriate traffics significantly. Recently we have implemented the dynamic keys in all forms available in the Portal to avoid massive automatic postings. However this regulation would reduce the interactivity among the visitors since the most popular media, that is the open forum, is open only for the registered members, although the threads are accessible for publics..

Optics and its related fields is a very active branch of science, also in Indonesia and the surrounding countries. So the Opto Portal has a great potential to be one of the most popular scientific portals in Indonesia. We would like to emphasize that there is a great facility on this Portal that is not yet widely used. The facility is the OPEN DATABASE which enables the contributors to open their (optics related) scientific databases for public. The database also accommodates partially free database, which enable the contributors to earn financial benefit from their valuable databases. No matter the database is completely or partially free to public, the contributors may use this generic database system for free.

The whole statistics to see (and to evaluate the progress of this Portal such as the monthly hits) are available online at http://www.opto.lipi.go.id/statistik/. Figure 3 shows the statistics' page of this Portal.

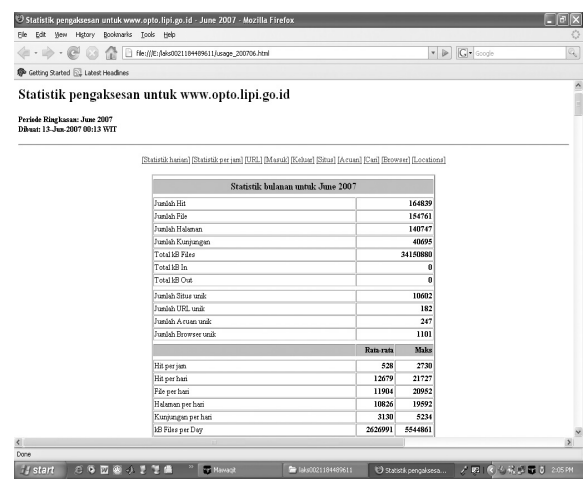

Figure 3. Statistics' page of the Portal.

## IV. CONCLUSIONS AND SUGGESTIONS

We have developed a scientific portal dedicated to the optical communities in Indonesia [5]. We realized that this Portal is still not popular and most of our aims are still not reached, in particular to pioneer the communities' online activities with real impacts and benefits for our society. In our opinion this Conference should be a good place to introduce and to popularize the existence of this Portal. We urge honored attendances, especially whom do not know yet this Portal, to start visiting it, utilizing the features offered in the Portal, and giving feedbacks to the Maintainers. After all, we should collaborate to properly utilize this Portal for the advantages to the optical communities in Indonesia and for the sake of our society at large.

For the future, as suggested by Martin Grötschel [4], we do hope that we can build partnerships with Scientific Libraries, Universities, Research Institutes, Database Providers, Document Delivery Services, Scientific Societies, Publishers, Software Houses, Data (Collecting) Centers, etc. for the benefit of Student (to access vast amount of materials), Employee (further training, lifelong learning), Teacher (reuse of high quality materials), Author (publishing cheap, fast, and widely), Publisher (open sources generate new chances), Business (more pofit from applying science), Citizen (contacting research more directly), Science (communicating with the public), and Society (free flow of information).

## ACKNOWLEDGMENT

We greatly appreciate generous supports from the LIPI Management and the Indosat Mega Media (IM2) for providing the infrastructures. We also thank the TGJ LIPI, the joint team for networks, for technical supports keeping this service available for public. Last but no least, we are very grateful to the contributors and visitors who already utilized this Portal.

## REFERENCES

[1] Chaonhan Youn et.al., "The GEON Portal: Scientific Portal Development for Research and Education", University of California, http://www.geongrid.org/.../abstracts/...
[2] -, "Web Portal – Wikipedia, the free encyclopedia", http://en.wikipedia.org/wiki/Web_portal
[3] David Morrison et al., "Building a Portal with Lotus Domino R5", IBM International Technical Support Organization, October 1999
[4] Martin Grötschel, "On the Road to Scientific Information Portals: Cooperative Digital Libraries. Remarks, Visions, Proposals", Universität Trier, 2001
[5] L.T. Handoko, Indonesian Copyright, No. B 268484 (2007).